# Disordered plasmonic system with dense copper nano-island morphology

*Tlek Tapani[1], Vincenzo Caligiuri[2,3], Yanqiu Zou[4], Andrea Griesi[2], Yurii P. Ivanov[2], Massimo Cuscunà[5], Gianluca Balestra[5], Haifeng Lin[1], Anastasiia Sapunova[2], Paolo Franceschini[6], Andrea Tognazzi[6], Costantino De Angelis[7], Giorgio Divitini[2], Riccardo Carzino[2], Hyunah Kwon[8], Peer Fischer[8], Roman Krahne[2], Nicolò Maccaferri[1] and Denis Garoli[2,9*]*

[1] Department of Physics, Umeå University, Linnaeus väg 24, 901 87 Umeå, Sweden
[2] Istituto Italiano di Tecnologia, Via Morego 30, Genova, 16163 Italy
[3] Dipartimento di Fisica, Università della Calabria, via P. Bucci 33b, 87036 Rende (CS), Italy.
[4] College of Optical Science and Engineering, Zhejiang University, Hangzhou 310027, China.
[5] Institute of Nanotechnology - CNR NANOTEC c/o Campus Ecotekne, Via Monteroni 73100 Lecce (Italy)
[6] Università degli studi di Palermo, Dipartimento di ingegneria. Viale delle scienze ed. 10, 90128, Palermo, Italia
[7] Università degli Studi di Brescia, Dipartimento di Ingegneria dell'Informazione, via Branze, 38 25123 Brescia, Italia
[8] Institute for Molecular Systems Engineering and Advanced Materials, Heidelberg University, 69120 Heidelberg, Germany; Max Planck Institute for Medical Research, 69120 Heidelberg, Germany
[9] Dipartimento di scienze e metodi dell'ingegneria, Università di Modena e Reggio Emilia, Via Amendola 2, 42122, Reggio Emilia, Italy

Email: denis.garoli@unimore.it

# Abstract

Dry synthesis is a highly versatile method for the fabrication of nanoporous metal films, since it enables easy and reproducible deposition of single or multi-layers of nanostructured materials that can find intriguing applications in plasmonics, photochemistry and photocatalysis, to name a few. Here, we extend the use of this methodology to the preparation of copper nano-islands that represent an affordable and versatile example of disordered plasmonic substrates. Although the island morphology is disordered, the high density of these nanostructures with large surface area results in a good homogeneity on a macroscale, which is beneficial for plasmonic applications such as bio-sensing and photo-catalysis. With cathodoluminescence and electron-energy-loss spectroscopies we confirm the nano-islands as sources of the local field enhancement and identify the plasmonic resonance bands in the visible and near-infrared spectral range. The decay dynamics of the plasmonic signal are slower in the nano-island as compared to bulk copper films, which can be rationalized by a reduced energy dissipation in the nano-island films. These properties make nano-islands films appealing as substrates for non-linear optics, for example in second-harmonic generation, for which we observe an efficiency comparable to silver thin films. Our study demonstrates a robust and lithography-free fabrication pathway to obtain nanostructured plasmonic copper substrates that represent a highly versatile low-cost alternative for future applications ranging from sensing to photochemistry and photocatalysis.

# Introduction

Random distributions of metallic nanostructures enable manipulation of light over a broad spectral range where the different geometries lead to light localization with different spatial confinement, from one to tens of nm, and to a landscape of spectrally diverse electromagnetic field enhancement. In addition, the irregular surfaces at the subwavelength scale produce an inhomogeneous local refractive index and large active surface area. Nanoporous metallic structures, for example, comprise abundant nanoscale gaps and sharp edges that interact with the incident light to generate a rich pattern of hot-spots.[1–3] In this context, we have recently demonstrated a dry synthesis method for the preparation of metallic nanoporous films which can find interesting applications in plasmonics.[4] Metallic films with a dense nanoisland morphology are another appealing platform for metamaterials and plasmonics, since the high density of sharp metallic structures and nanogaps are exposed on the surface, and therefore available for electrochemical and optical sensing [5], photochemical catalysis[6], and advanced energy technologies[7].

Among metals, copper is highly interesting due to its plasmonic properties and low-cost, and therefore copper nanostructures caught attention in a variety of areas such as catalysis, sensors and batteries.[8–10] Consequently, copper nano-islands (Cu-NIs) are promising for a range of applications. However, fabricating pure, homogeneous Cu-NIs remains a challenging task in nanomaterial synthesis. One of the most commonly employed techniques is thermal dewetting [8,11–13]. Although thermal dewetting is straightforward, it offers limited control over the uniformity, thickness and distribution of the resulting nano-islands. The spontaneous nature of this process often leads to irregularities in size and spacing, which can affect the material's overall properties and performance. Another method used to fabricate Cu-NIs is laser ablation,[14,15] where a high-energy laser hits a copper target, generating a plasma plume. The copper atoms within this plume then deposit onto a substrate, forming nano-islands. This technique allows for a higher degree of precision in controlling the size and uniformity of the nano-islands. However, achieving optimal results with laser ablation requires meticulous adjustment of several parameters and careful control of substrate conditions. This makes the process complex and labour-intensive, often requiring significant trial and error steps to finely tune the desired outcomes. Nanoparticle self-assembly is another approach,[16,17] in which pre-synthesized copper nanoparticles are deposited onto a substrate, where they self-assemble into nano-islands. Self-assembly offers the advantage that well-defined nanoparticles can be used, however, control of the assembly process can be difficult, leading to a large variation in the distribution and morphology of the nano-islands. This variability can be a significant drawback in applications that require highly uniform and homogeneously distributed nanostructures.

In this work, we modify our technique for the fabrication of nanoporous metals [4] to obtain films of Cu-NIs by a simple and reproducible method. The process involves the plasma treatment of a bilayer consisting of copper deposited by physical vapor deposition (PVD) on top of a sacrificial PMMA layer. This approach offers the flexibility to tailor the material's properties by varying the deposition parameters and plasma conditions, while also providing a relatively simple, lithography-free, large-area fabrication method with low-cost processes. Furthermore, our method, not only, enables the preparation of pure Cu nanostructures, but it can be easily extended to produce CuO nano-islands/nanostructures that are widely used in applications such as catalysis[18,19] energy storage[20], sensors[21,22], and antimicrobial agents[23].

Here we performed a detailed characterization of morphological, optical and electronic properties of the Cu-NIs by using a unique combination of experimental techniques. Electron microscopy and spectroscopic ellipsometry were first used to characterize the morphology and the optical response of the system, then, with electron energy loss spectroscopy (EELS) and cathodoluminescence (CL) we investigated the spatial hot spot distribution and spectral bands of the plasmonic resonances. Moreover, X-ray Photoelectron and Surface Enhanced Raman Spectroscopies (XPS and SERS) were used to verify the level of oxidation and the potential enhanced spectroscopy in the produced structures. Finally, electron dynamics and nonlinear optical properties were evaluated by pump-probe spectroscopy and second harmonic generation (SHG).

# Results

The dry-synthesis approach [4,24] is based on the plasma treatment of a dense layer of nanoparticles deposited on a sacrificial thin PMMA polymer film. The metal layer is deposited by e-beam evaporation at an oblique angle between 70° and 80° on top of the sacrificial layer, which is then removed by plasma etching. The etching of the sacrificial PMMA layer was done using $N_2$ plasma to avoid oxidation of the copper. This has been proved by means of XPS, comparing the preparation of the NIs using $N_2$ and $O_2$ plasma. As reported in Supporting Information – Note#1, while the use of $N_2$ ensures the preparation of Cu nanostructures with very low oxygen content (also after two days in air), if $O_2$ plasma is used, CuO nanostructures are obtained, as expected. A key parameter in the preparation of Cu-NIs is the deposition rate. With a Cu deposition rate of 0.1 nm/sec a dense array of Cu nanoparticles is obtained on top of the PMMA, and after the plasma etching these particles are clustered in Cu-NIs (see Supporting Information – Fig. S2 and S3). The deposition and etching steps can be repeated to fabricate thicker and denser films with nano-island morphology. We note that by using higher deposition rates (>3 nm/sec) a denser Cu film is formed prior to the etching, and then the plasma etching produces a nanoporous morphology (see SI – Fig. S6). Fig. 1 shows SEM images of the Cu-NIs films obtained after two fabrication steps (two repeated layers[4,24]) based on the low deposition rate that evidence the dense nano-island morphology. We observe several differently sized features in the nanoisland film: (i) sharp tips of the islands with less few nm diameters and several tens of nm length, (ii) island size of several hundred nanometers, and (iii) gaps in between the islands of submicron dimensions. The combination of these features provides an excellent example of a randomly disordered plasmonic system that can be expected to sustain multiple resonances at different energies. The dielectric function and complex refractive index of the films measured by spectroscopic ellipsometry are reported in Supporting Information – Note#2.

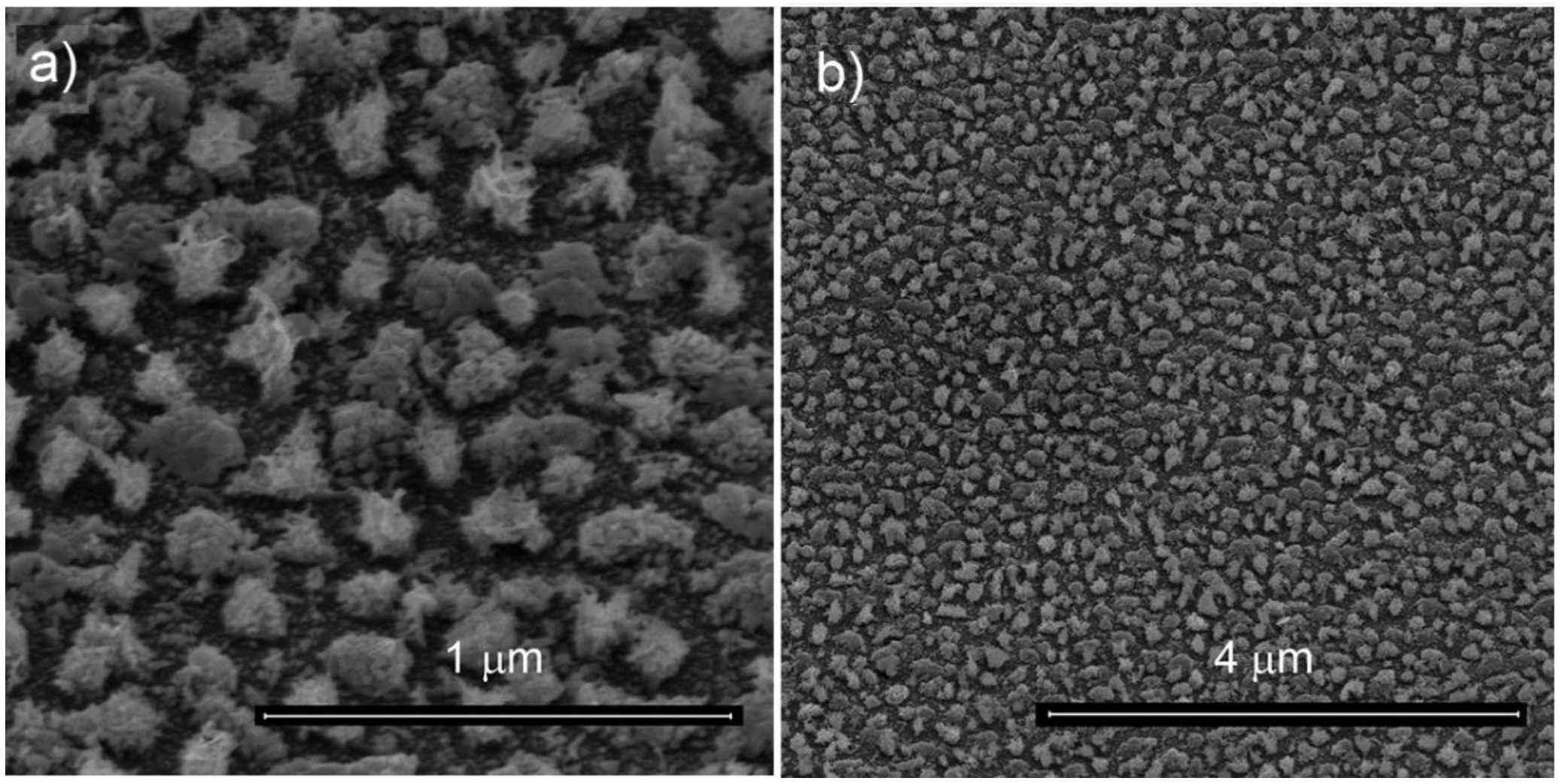

**Figure 1**. Tilted view SEM images of the Cu-NIs films at different magnification (obtained with two evaporation+plasma etching steps).

The dense, random network of Cu-NIs, with varying sizes, is expected to exhibit localized surface plasmon resonances (LSPRs) at different energies. We used CL and EELS as complementary tools to investigate the energy bands of the LSPR, and to map their spatial distribution. With CL the bright, emissive modes of the substrate can be investigated. On the contrary, EELS is sensitive to the dark and/or non-radiative modes, and therefore gives complementary information on the spectral properties of the plasmon resonances. CL can be directly performed on films as displayed in Figure 1, while EELS samples need to be thin enough to limit interaction with the electron beam, so dedicated specimens were produced. We therefore fabricated a Cu-NIs film onto a thin silicon nitride membrane (30 nm thickness), to ensure a minimal background to the signal.

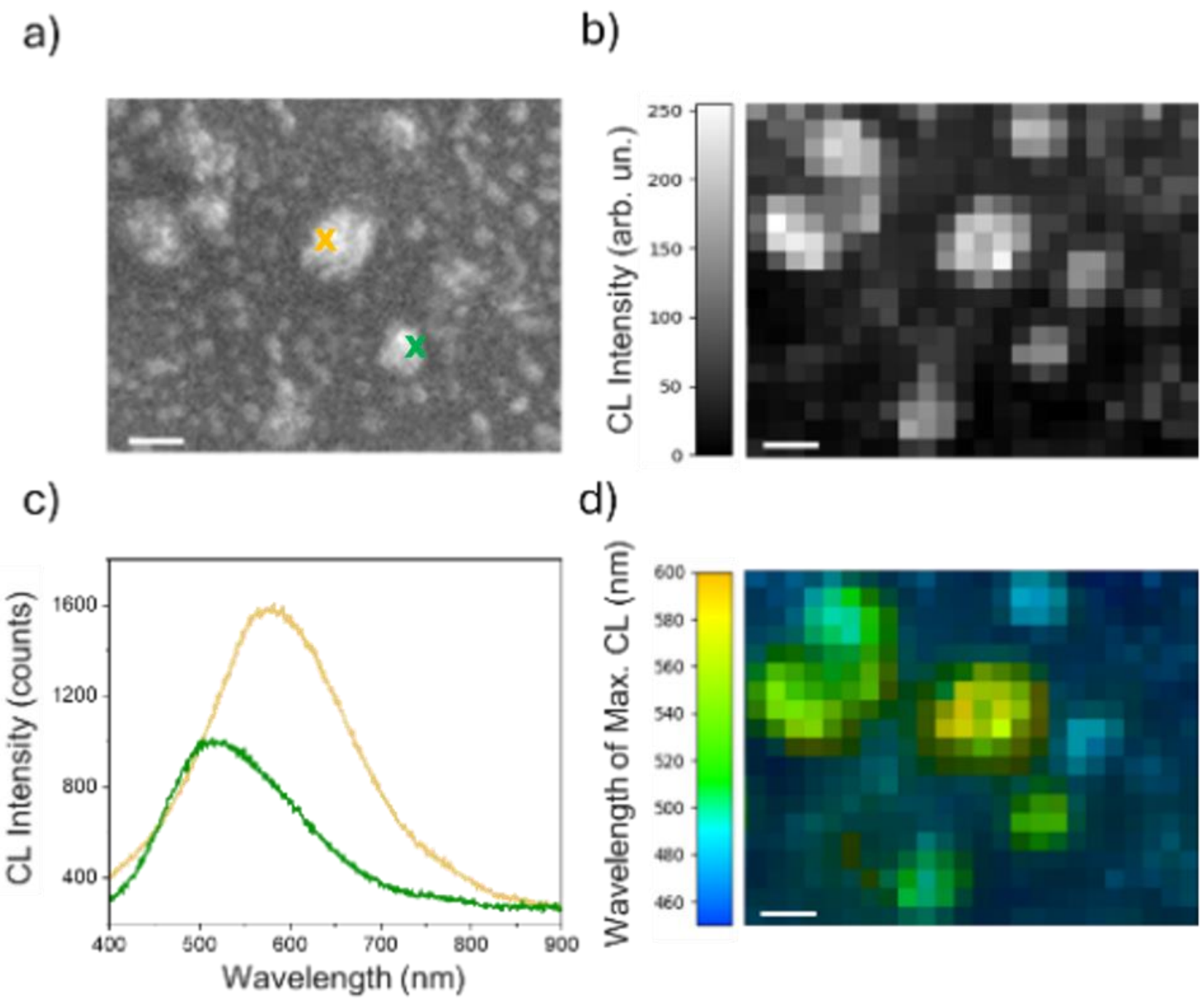

**Figure 2.** (a) SEM image of the copper nano-island film investigated by cathodoluminescence (CL). (b) Panchromatic CL map of the region in a) spanning a wavelength range of 400-900 nm. (c) Local CL spectra extracted from two positions indicated by crosses of the corresponding color in a). (d) Color-coded map showing the dominant wavelength of CL emission for each pixel in the panchromatic map in b). Scale bars in b) and d): 100 nm.

The CL of a typical region of a Cu-NIs film is reported in Figure 2. The comparison of the morphology (SEM image in Figure 2a) with the CL intensity (Figure 2b, and the panchromatic map, Figure 2d) confirms the nano-islands as a source of LSPRs, and allows to correlate larger islands with more red-shifted CL peaks, and smaller islands with more blue-shifted CL peaks (see yellow and green spots and spectra in Figure 2a,c, respectively). The CL emission over subwavelength distances arises from the excitation of surface plasmon (SP) modes in the Cu-NIs network that generate numerous localized emission spots. The panchromatic map in Figure 2d displays the distribution of these SP modes, and the dominant wavelength of CL emission for each pixel. Here, the panchromatic optical response represents the total light intensity captured during the dwell time of the electron beam mapped onto each pixel. This map highlights variations in SP mode energies across neighboring regions, reflecting the interplay of metal clusters and empty spaces within the copper network. The main CL is in the range from 530 to 590 nm, with larger Cu clusters exhibiting a red-shifted emission compared to smaller ones.

The results of the EELS analysis are depicted in Figure 3 and Supporting Information – Note #3. Figure 3a reports the annular dark field (ADF) image. Note that although the data for the image is acquired in dark field mode, due to the low acceleration voltage of 60 keV the Cu-NIs appear dark caused by their strong electron scattering, while the bright areas represent the gaps.  EELS investigates local electronic properties by quantifying the energy loss of electrons interacting with the sample using a nm-sized electron probe. Such analysis was restricted to the gap areas, observing the energy losses linked to excitations that can be activated in so-called “aloof mode”, as routinely done in the literature for metal nanoparticles. We identified the relevant spectral regions by applying a threshold to the intensity of the zero-loss peak (ZLP), which is directly correlated with the local thickness. In the spectrum, the ZLP comprises electrons that interact elastically with the sample, resulting in no measurable energy loss. The tail of the ZLP was removed using a power law fit. We then employed Non-negative Matrix Factorization (NMF)[25–27], a multivariate analysis approach that detects patterns by analysing local correlations and generating components based on statistical criteria, with minimal operator input. This approach minimizes human bias and is particularly effective for extracting meaningful features from hyperspectral data[28,29]. The factors present a set of maxima that begin at ~0.33 eV, followed by peaks at higher multiples of the lowest energy, which we attribute to the higher order harmonics of surface plasmon excitation.[30] Interestingly, the difference between excitations (~0.3 eV) is equal to the difference between the two emission peaks found using CL (Figure 2c) in the visible range (570nm -> 2.17eV and 500nm -> 2.47eV).

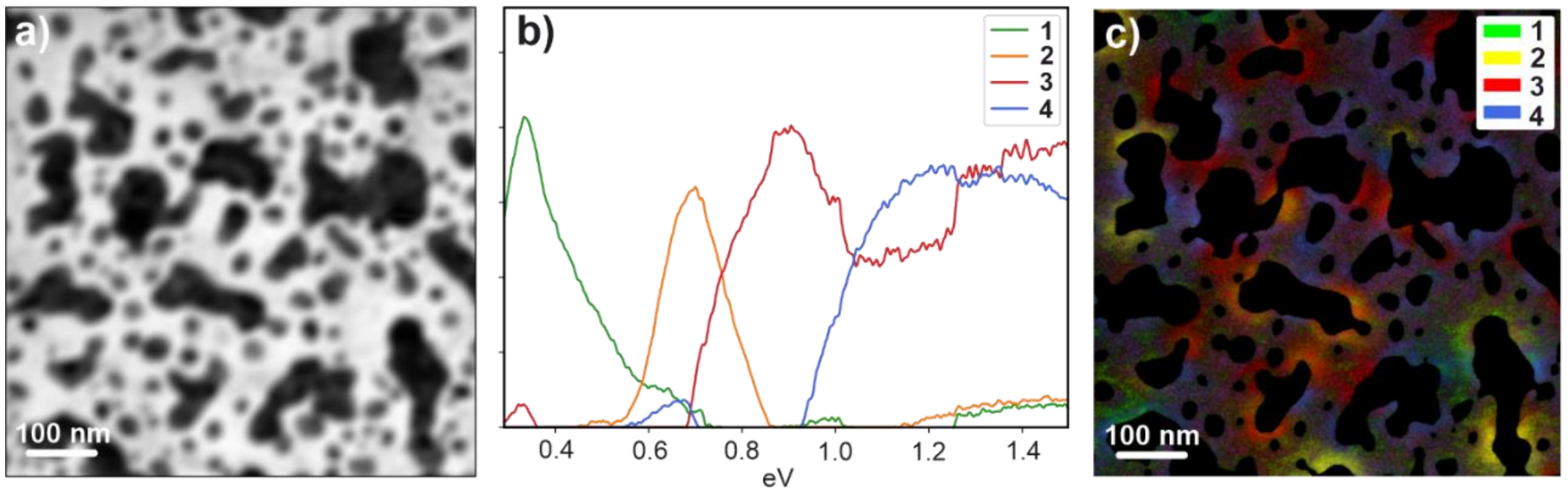

**Figure 3.** EELS measurements of the Cu-NIs film. (a) Reference ADF (Annular Dark Field) image – note that contrast appears reversed due to the total thickness of the film and membrane. (b) Energy profiles of NMF components (c) Spatial distribution of the NMF components.

The signal/background ratio and the sharpness of the EELS features linked to plasmons decrease at higher orders, meaning that detection of excitations in the visible range becomes particularly challenging: in this case, the NMF algorithm does not produce components at higher energy losses.

To spatially localize the four energy loss mechanisms identified by NMF, we created a color map using the loadings associated with each NMF component (Figure 3c), where the color intensity of each pixel reflects the weight of the corresponding component at a given position in the scan. To display the spatial distribution of the various modes and correlate them to the local geometry, we selected a representative area and produced spatial maps reporting the total intensities within a defined spectral window (see Figure S10 in the SI). Extending the analysis to an energy interval centered at 2.17 eV and a width of 0.3 eV, we also find a weak EELS signature for a higher excitation mode, in agreement with the CL observations.

While EELS and CL allowed us to access the nanoscale scattering and absorption properties of the Cu-NIs, pump-probe transient transmission measurements were employed to explore the ultrafast dynamics of photoexcited states, and to provide a deeper understanding of electronic properties responsible for the plasmonic behaviour. This technique allows to observe the temporal evolution of the LSPRs, which can significantly impact its coupling to other processes, such as chemical reactions in photocatalysis[31], surface-enhanced Raman scattering[32] and exciton dynamics in light-harvesting complexes[33]. In Figure 4a,b, we report the transient transmission of both the continuous and Cu-NIs films of thickness of ~25nm as function of wavelength (x-axis) and time delay between the pump-probe pulses (Δt, y-axis). In Figure 4a, a pronounced negative signal appears around 577 nm, corresponding to a photon energy of 2.15 eV, which can be related to a transient induced absorption due to hot carrier excitation. Such transient absorption can be explained by considering the excitation of 3d to 4p inter-band transition in Cu[34], as previously reported for similar nanostructured/porous gold films[35]. Even though the pump pulse, with center photon energy of 1.5 eV, has energy below the interband transition threshold, it creates a population of hot electrons that has sufficient energy for interband transitions. Thus, empty states that can be occupied by electrons excited by the probe pulse at lower energies are created, and this effect amplifies absorption, resulting in a reduction of the probe signal around 577 nm. The same behavior is observed for the NIs film below 600 nm, but it is blue-shifted due to the LSPR from the Cu-NIs that is located between 530 and 590 nm. Here the LSPR enhances the light-matter interaction and amplifies the observed effect.

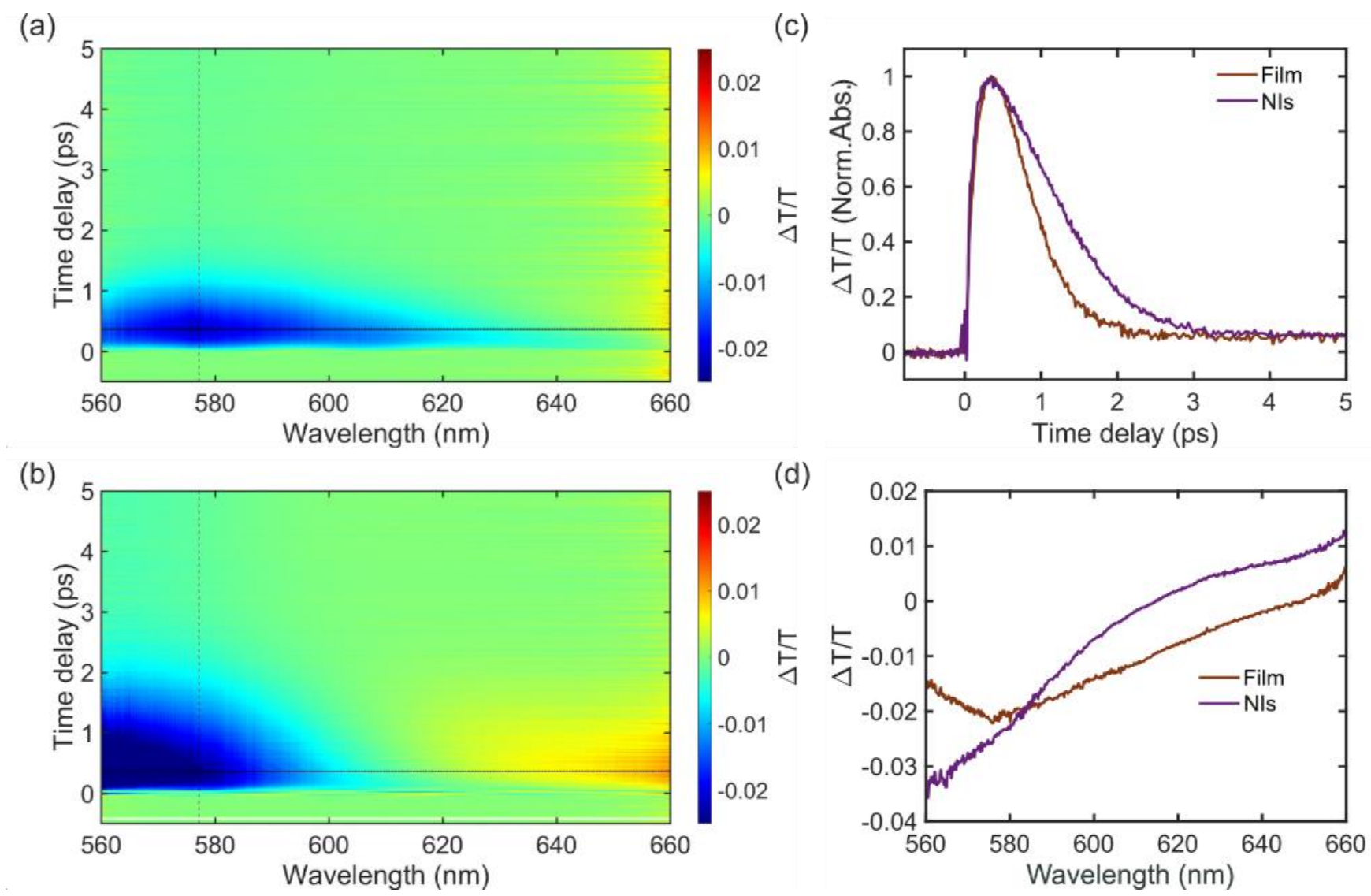

**Figure 4.** Ultrafast charge dynamics of an unstructured (a) and nano-island (b) Cu film. (a/b) transient transmission color map as function of wavelength and time delay of the film/NIs sample. (c) Normalized charge dynamics comparison of film and NIs at 577 nm wavelength (vertical dashed line in color map (a,b)). (d) Normalized transient transmission amplitude of a continuous film and NIs at fixed time delay of 0.37 ps (horizontal line in color maps (a,b)).

In Fig 4c, the charge dynamics extracted from the color maps (at the wavelength of 577 nm marked by the vertical black dotted lines in Figure 4a,b) are depicted. The signal amplitude has been normalized to facilitate a clear comparison of the relaxation dynamics (exponential decay after excitation). Following the pump excitation, a simultaneous alteration of the transient transmission signal can be noticed, indicating similar initial charge excitation dynamics. However, this is followed by clearly different recovery processes. We evaluated the decay of the transient signal by exponential fitting, obtaining a rapid recovery with life time of around 600 fs for the continuous film, which is half of that of the NIs sample of around 1200 fs. Consistent with ref [35], the isolated geometry of the NIs decreases their thermal capacitance, reducing their thermalization and eventually leading to a higher electron temperature. In Figure 4d, the transient transmission amplitudes of both the copper NIs and continuous film are plotted at 370 fs following pump pulse excitation. This specific time delay is selected because it corresponds to the point where the transmission variation signal reaches its maximum (see Figure 4c). In Figure 4d, the transient transmission amplitude of the NIs sample shows a greater modification compared to the film. In the blue region of the detection range (around 560 nm), a larger negative ΔT/T value suggests a higher photon absorption within the Cu-NIs sample. Conversely, in the red region of the probe spectrum (around 660 nm) where pump mainly excites electron in the conduction band (promoting mainly intra-band transitions), a higher positive ΔT/T value is observed, indicating a much higher electron temperature in the NIs sample[37] and reduced energy dissipation.[38]

Due to the presence of multiple hotspots in a broad wavelength range and at sub-wavelength distances, disordered plasmonic systems such as the Cu-NIs film can be expected to feature strong nonlinear optical properties, which can be exploited for applications ranging from sensing to image recognition.[39–41] Therefore, we measured the SHG signal from our samples using a nonlinear microscope working in reflection configuration. Here a linearly polarized laser beam at fundamental frequency (FF) $\lambda_{FF}$=880 nm impinges on the sample surface at normal incidence after being focused to a spot size (waist) $w_0$=4.5 μm by a 20x-objective. The back-scattered nonlinear signal is collected by the same objective and its intensity in the visible spectral range (400-500 nm) is measured by a single photon avalanche detector (see Methods for more details). Given the uniform morphology on a macroscale of the Cu-NIs films, we performed space-resolved nonlinear generation measurements by scanning an arbitrary square region (100 x 100 $\mu m^2$) on the sample surface and report the power-dependent measurements in Figure 5. By fitting a power-law profile (solid black line) to the experimental data, the analysis provides a value of the exponent of 1.93±0.08, which suggests a second-order nature of the nonlinear process. Such result and the presence of spectral filters in the detection line (see Methods) allow to ascribe the nonlinear signal to SHG process. The second-harmonic conversion efficiency is $\eta_{SH}=P_{SH}/P_{FF}=4\times10^{-13}$, with $P_{FF}$ ($P_{SH}$) being the average power of the fundamental frequency (second harmonic frequency) radiation impinging on (back-scattered from) the sample. Such value is larger with respect to efficiencies obtained for aluminum surfaces [42] and comparable with efficiencies obtained in silver thin films under oblique illumination [43]. Moreover, it is only one order of magnitude lower with respect to other nanostructured metallic antenna arrays [44], highlighting the Cu-NIs compound as a promising plasmonic material platform. The comparison between two space-resolved SHG scanning maps obtained from the same sample region with two orthogonal directions of the FF beam polarization shows that the SHG signal intensity does not depend on the FF beam polarization direction (see Supporting Information -Note#5).

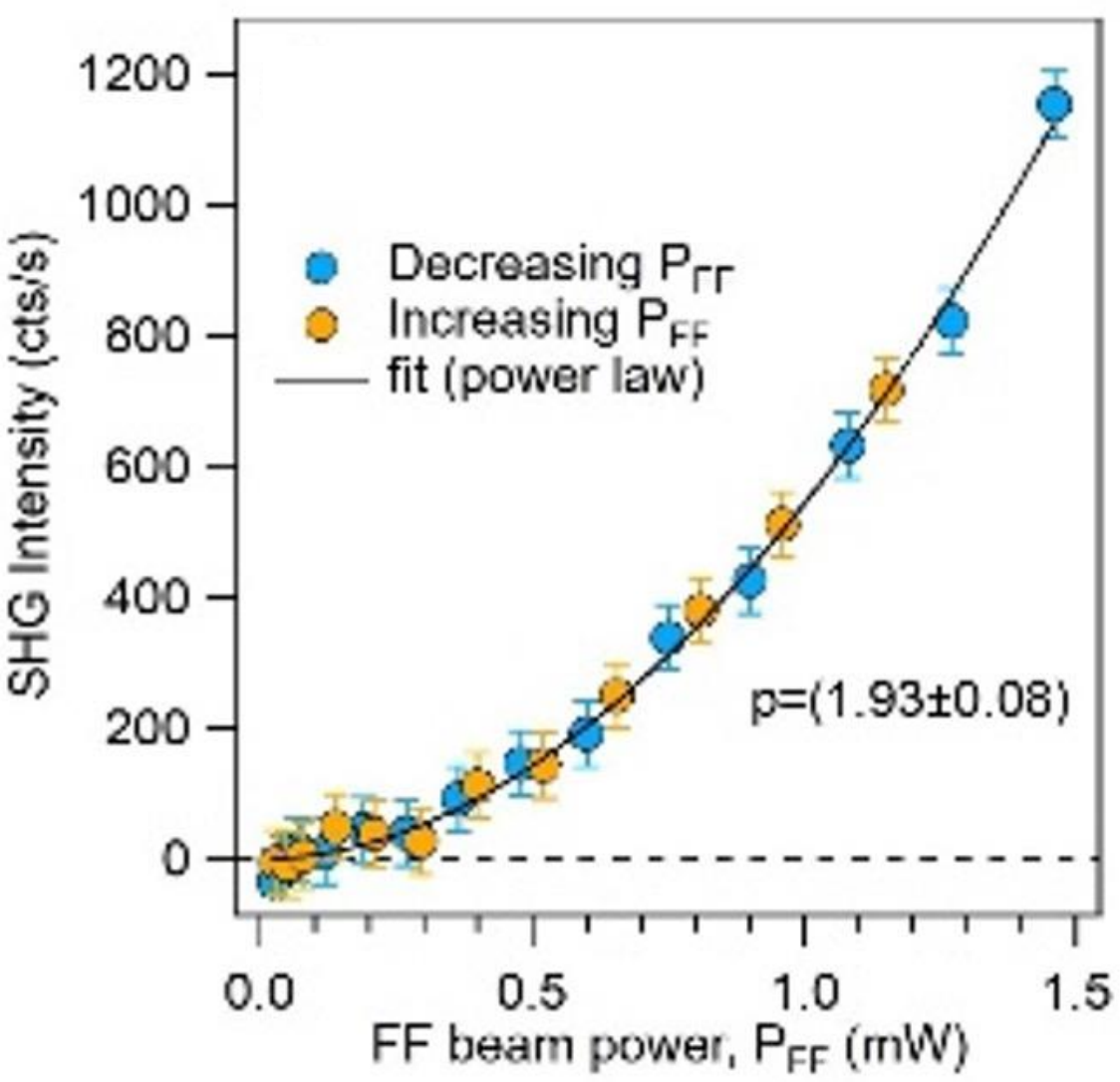

**Figure 5.** SHG response of Cu-NIs. Power-dependent measurement of the SHG signal from the Cu-NIs: experimental data (markers) while decreasing (light blue) and increasing (yellow) FF beam power and power-law theoretical profile (black solid line).

## Conclusion

In this study, we successfully extended the dry synthesis method to prepare cost-effective Cu-Nis films that represent an interesting and affordable disordered plasmonic material with diverse potential applications ranging from sensing (as demonstrated for example in the preliminary SERS experiments reported in Supporting Information – Note#6) to photocatalysis. This fabrication process is universal, applicable to different metals, and can be performed in multiple steps, allowing for the creation of multicomponent films. This opens the door to more complex and versatile applications based on various materials and optical processes. The optical characterisation with CL and EELS of the Cu-NIs films demonstrated their robustness, high density, and multiple bands nature. Ultrafast spectroscopy revealed that the relaxation time in the Cu-NIs is slower compared to their bulk counterparts, which is advantageous in photocatalysis where the hot electron dynamics can be tailored to control ultrafast chemical processes. Nonlinear optical measurements further showcase that these Cu-NIs can support SHG processes without the need for high-power laser sources, making them highly suitable for practical applications in nonlinear optics, such as frequency conversion, optical switching, and signal processing in photonic devices.

# Methods and materials

*Samples preparation:*

The samples preparation is based on the original methods proposed in ref.[24]. In brief, Poly(methyl methacrylate) (PMMA) was spin-coated on a suitable substrate (Silicon for SEM and elipsometry, thin SiN membrane for EELS and CL, fused silica for ultrafast dynamics) at 4000 rpm for 2 min. Each metal (>99.99% purity) was evaporated by e-beam on a PMMA thin film at room temperature with an oblique angle of 80°, a rate of 0.1 nm/s, a target thickness of 12 nm was used in all the cases. The deposited Cu film was plasma treated in $N_2$ with 200 W till the whole PMMA layer was removed. The thickness of the final film was tuned repeating the process multiple times.

*Spectroscopic Elipsometry:*

Spectroscopic ellipsometry was carried out by a M2000 apparatus by Woollam. Spectroscopic analysis was carried out at three different angles (50°, 60°, 70°). The fitting procedure was carried out by starting with a point-by-point fit of the pseudo-dielectric permittivity obtained directly from the measured ellipsometrical angles. The obtained imaginary dielectric permittivity was then fitted as a convolution of oscillators, as described in the main manuscript, to obtain a Kramers-Kronig-consistent analytic expression for the effective dielectric permittivity of the NPMs. The real parts of the dielectric permittivities have been obtained by applying Kramers-Kronig relations to the imaginary parts.

*EELS measurements:*

The films were directly fabricated on silicon nitride chips with a 30 nm membrane. Analysis was carried out at 60 kV acceleration voltage in a ThermoFisher Spectra300 S/TEM equipped with a monochromator with “UltiMono”. The energy resolution of the spectra was ∼40 meV. The data were analysed using Hyperspy, a python-based data analysis toolkit.

*Cathodoluminescence measurements:*

Nanostructured Cu films were fabricated on a 100 nm thick Si3N4 membrane to significantly minimize emission from bulky substrates during cathodoluminescence (CL) investigation. CL analysis was performed at room temperature by using a Zeiss Merlin scanning electron microscope (SEM) equipped with a high-performance CL imaging system (SPARC from Delmic). CL was spectrally resolved in the range of 400-900 nm with an “Andor Kimera 193i” spectrometer with a focal length of 193 mm and a grating of 300 gr/mm.

The photon emission was captured by an “Andor Newton DU920P-BEX2DD” CCD camera with a maximum quantum efficiency of 90%. The electron beam operated at an acceleration voltage of 30 kV and an emission current of 10 nA. The panchromatic CL map consisted of 24 × 19 pixels (pixel size: 32 nm). The focused electron beam was scanned across the specimen, dwelling for 10 s (integration time) on each pixel to acquire the CL spectra. The intensity in each pixel of the panchromatic map corresponds to the integrated intensity of the detected light during the integration time.

*Pump-probe experiments*

Pump-probe optical spectroscopy experiments utilized sub-15 fs light pulses to investigate carrier dynamics in both Cu thin film and Cu NIs samples. The sketch of the setup is shown in Supporting Information note-4. We employed a Yb:KGW amplifier laser with an operational wavelength of 1030nm, a repetition rate of 50 KHz, and an average output power of 20 W, driving two separate optical parametric amplifiers (OPAs). The outputs from these OPAs acted as the pump-probe pulses for the measurements, with center wavelengths of 850nm (1.5 eV) and 600nm (2 eV), respectively, as illustrated in Supplementary Figure S1b. The time delay between the pump-probe was precisely controlled with 1 fs accuracy using a high-precision linear stage (model: PI L1556ASD00) positioned along the pump beam path. Our experimental focus involved measuring transient transmission, which refers to the pump-induced changes in transmission, as defined by

$$\Delta T/T=(T_t-T_0)/T_0$$

Where $T_t$ and $T_0$ respectively represent the transmission of the system following excitation and at a particular time delay between the pump-probe pulses, and in the ground state conditions where the pump pulse does not induce excitation in the system.

*SHG Measurements*

SHG measurements have been performed by a home-built nonlinear microscope similar to the one described in [45]. The system (see Figure S12 of the Supplementary Information note-5) is based on a femto-second laser (Monaco by Coherent) generating 300 fs temporal duration, 40 µJ energy, 1035 nm wavelength pulses at 500 MHz repetition rate. The laser output feeds an optical parametric amplifier (OPA, Opera-F by Coherent) producing a broadband radiation at around 880 nm wavelength, whose spectral width is then reduced by a band-pass filter (FBH880-10 by Thorlabs). The obtained radiation has been employed as the fundamental frequency (FF) beam of the nonlinear experiment. The FF beam intensity is tuned by an attenuation system combining a half-wave plate (AHWP05M-980 by Thorlabs) and a polarizer (GL10 by Thorlabs). Before reaching the sample, the FF beam polarization (linear) is controlled by an additional half-wave plate. A 20x objective (LMPLFLN20X by Olympus) focuses the FF beam onto the sample surface with a spot size of 4.5 μm (radius at $e^{-2}$) and collect the generated nonlinear signal. The back-scattered SH signal is separated from the FF by a dichroic long-pass filter (69-891 by Edmund Optics), and its intensity is measured by a single photon avalanche detector (SPAD, PD-50-0TD by MPD). A set of short-pass filters (FESH0500, FESH0550, and FGB18S by Thorlabs) are placed before the SPAD to remove any FF beam residuals.

The sample surface and the position of the FF beam spot on the sample surface can be visualized by a (linear) optical microscope enclosed in the nonlinear setup. Here, the light from a LED source illuminates the sample surface and the corresponding image is obtained on a CCD camera (CS165MU/M by Thorlabs) by employing a suitable lens system which includes the 20x objective.

*X-Ray Photoelectron Spectroscopy Measurements*

Surface chemical composition and the presence of specific functional groups were investigated by means of X-ray Photoelectron spectroscopy (XPS, Kratos, Axis UltraDLD). Both survey spectra and high resolution ones were acquired, using a monochromatic Al kα source operated at 20 mA and 15 kV. Survey spectra were acquired at pass energy of 160 eV, energy step of 1 eV and over an analysis area of 300 × 700 μm. High resolution spectra were acquired on the same area, at pass energy of 20 eV and with an energy step of 0.1 eV. The Kratos charge neutralizer system was used on all samples; binding energy scale calibration was performed by setting the position of the main C 1 s component, C-C bonds, at 285 eV for PMMA and 284.8 for the Adventitious Carbon. Casa XPS software were used to analyze the experimental data.

*Surface Enhanced Raman Spectroscopy Measurements*

SERS measurements were performed using a Horiba LabRAM HR Evolution Raman spectrometer (Horiba Jobin Yvon, Kyoto, Japan) with a 50× long-focal-length objective (NA = 0.75). Rhodamine 6G(R6G) with varying concentrations were measured using a 633 nm laser at 5% power (spot diameter ~1.03 μm) and a 532 nm laser at 5% power (spot diameter ~0.86 μm). Spectra were acquired with a 600 grooves/mm grating, 10 s exposure time, and one accumulation. For mapping measurements, a 5×5 grid (total 25 points) with a 5 μm step size was performed, covering a total area of 20×20 $\mu m^2$. The same laser and acquisition parameters were maintained for each measurement point. The spectral resolution was approximately 3–4 $cm^{-1}$ and 2–3 $cm^{-1}$ for different laser wavelengths, respectively.

## Acknowledgements

The authors thank the European Union under the Horizon 2020 Program, FET-Open: DNA-FAIRYLIGHTS, Grant Agreement 964995 and the HORIZON-Pathfinder-Open: 3D-BRICKS, grant Agreement 101099125; DYNAMO, grant Agreement 101072818. V. C. thanks the research project “Componenti Optoelettronici Biodegradabili ed Eco-Sostenibili verso la nano-fotonica "green”” (D.M. n. 1062, 10.08.2021, PON “Ricerca e Innovazione” 2014-2020), contract identification code 1062_R17_GREEN. TT, HL and NM acknowledge support from the Swedish Research Council (grant no. 2021-05784), Kempestiftelserna (grant no. JCK-3122), the Knut and Alice Wallenberg Foundation (grant no. KAW 2023.0089), the Wenner-Gren Foundations (grant no. UPD2022-0074) and the European Innovation Council (grant no. 101046920 ‘iSenseDNA’).